# SuperSkillsStack: Agency, Domain Knowledge, Imagination, and Taste in Human-AI Design Education

**Qian Huang**
Research Fellow, Lee Kuan Yew Center for Innovative Cities, Singapore University of Technology and Design

**King Wang Poon**
Director, Lee Kuan Yew Center for Innovative Cities; Chief Strategy and Design AI Officer, Singapore University of Technology and Design

**Abstract**

This study examines how students integrate generative artificial intelligence (AI) into design projects through the lens of the SuperSkillsStack framework, which identifies four key human competencies for effective human-AI collaboration: Agency, Domain Knowledge, Imagination, and Taste. As generative AI increasingly transforms creative practice, design education must consider how human capabilities are cultivated alongside technological tools. Using qualitative thematic analysis, this study analyzes reflective writings from 80 student design teams participating in a human-centered design course. The findings show that students primarily used AI during the early stages of the design process, including brainstorming, information synthesis, and problem framing. However, students consistently relied on human judgment to interpret contextual information, validate AI-generated outputs, and refine design solutions. Domain knowledge derived from field observations enabled students to detect inaccuracies in AI suggestions, while taste played a critical role in evaluating and selecting meaningful ideas. The results suggest that generative AI functions primarily as a cognitive accelerator rather than a replacement for human creativity. The study highlights the importance of cultivating higher-order human capabilities to support effective human-AI collaboration in design education.

**Introduction**

The rapid advancement of generative artificial intelligence (AI) has begun to reshape creative practice, knowledge work, and education. In design disciplines particularly, AI systems are increasingly used to generate ideas, synthesize research insights, and produce visual artifacts. While these technologies offer unprecedented efficiency and access to computational creativity, they also raise important questions about the future of human learning and expertise.

Design education has traditionally emphasized the development of human-centered capabilities such as observation, empathy, creative problem solving, and critical judgement. These capabilities enable designers to navigate complex social and environmental contexts while producing meaningful solutions. However, the growing accessibility of generative AI tools introduces new dynamics into the design process. Students now have access to systems capable of generating design suggestions, visual concepts, and analytical summaries within seconds.

This transformation has led educators and researchers to ask a fundamental question: How can design education integrate AI tools while still cultivating the uniquely human capabilities that underpin meaningful design practice?

Recent work in AI-assisted design pedagogy suggests that the integration of generative AI into learning environments requires a shift in educational focus toward higher-order cognitive capabilities rather than purely technical skills (Khan et al., 2025; Willems et al., 2024). Instead of replacing human creativity, AI tools can serve as collaborators that support exploration, reflection, and iteration in design processes. Research within engineering and design education has increasingly examined how generative AI tools can reshape learning environments, enabling students to engage in experimentation, reflection, and iterative design processes (Huang et al., 2025; Willems, Khan, et al., 2025).

Within this emerging perspective, designers must develop capabilities that allow them to critically interpret AI outputs while maintaining responsibility for contextual reasoning and creative judgement. Huang, Willems, and Poon (2025) describe this shift as part of a broader transformation in design education, where students must learn to orchestrate human–AI collaboration rather than rely solely on individual creative production.

To conceptualize these human capabilities, recent work proposes the SuperSkillsStack framework, which identifies four interrelated competencies necessary for effective collaboration with AI systems: Agency, Domain Knowledge, Imagination, and Taste (Huang et al., 2025).

These capabilities represent a layered cognitive framework enabling individuals to strategically deploy AI tools while maintaining control over interpretation and design decision-making.

Despite the growing interest in AI-assisted design workflows, empirical research examining how students actually integrate AI into design projects remains limited. In particular, little is known about how students balance AI-generated insights with human observation, contextual knowledge, and evaluative judgement.

This study addresses this gap by analyzing reflective writings from student design teams who integrated AI tools into their design workflows. Using the SuperSkillsStack framework as an analytical lens, the study examines how students exercised agency, applied domain knowledge, engaged imagination, and demonstrated taste when working with generative AI systems.

The research addresses the following questions:

1. How do students demonstrate agency when deciding how AI tools should be used in design projects?
2. How does domain knowledge influence their ability to interpret and evaluate AI-generated outputs?
3. In what ways does AI support or constrain imagination during the ideation process?
4. How do students exercise taste when judging the quality and relevance of AI-generated suggestions?

By examining these questions, the study contributes to ongoing discussions about how design education can evolve in response to generative AI technologies while continuing to cultivate human-centered capabilities.

**Literature Review**

**1) Generative Artificial Intelligence and Creative Practice**

The rapid advancement of generative artificial intelligence (AI) has significantly altered the landscape of creative practice and knowledge work. Systems based on large language models (LLMs) and generative visual models are capable of producing text, images, and design concepts from natural language prompts. These technologies are increasingly integrated into workflows across domains such as product design, architecture, interaction design, and visual communication.

Recent research suggests that generative AI systems can enhance creative processes by expanding the range of possible ideas explored during early design stages. Generative systems can produce multiple conceptual variations quickly, enabling designers to engage in rapid ideation and exploration (Elgammal et al., 2017; Willems et al., 2024). Such systems are particularly useful during the divergent phases of design thinking, where generating a wide variety of possibilities is essential to discovering innovative solutions (Brown, 2008).

Recent studies have also begun examining how humans perceive creativity produced by AI systems. Experimental research comparing human and AI-generated creative outputs suggests that while AI can produce novel artifacts, human judgement remains central in evaluating originality and usefulness (Deng et al., 2024). These findings reinforce the view that AI systems can support creative exploration but do not replace the evaluative role of human designers.

At the same time, scholars emphasize that generative AI does not possess contextual understanding in the same way human designers do. Because these systems rely on statistical patterns extracted from large datasets, they may produce outputs that appear plausible but lack grounding in specific social, cultural, or spatial contexts (Boden, 2016;

Floridi & Chiriatti, 2020). As a result, AI-generated design suggestions often require human interpretation and refinement before they can be applied to real-world problems.

These limitations have led researchers to frame generative AI not as a replacement for human creativity but as a complementary system that augments human creative processes (Daugherty & Wilson, 2018). Within this perspective, the role of the human designer shifts from producing ideas independently to orchestrating interactions between human insight and computational generation.

2) **Human–AI Collaboration in Design Processes**

The concept of human–AI collaboration has become central to discussions about the future of creative work. Rather than viewing AI as an autonomous creator, scholars increasingly describe AI systems as collaborators that contribute specific capabilities within hybrid creative systems (Amershi et al., 2019; Willems et al., 2024).

AI systems demonstrate particular strengths in tasks involving: large-scale information processing; pattern recognition across datasets; rapid generation of multiple variations.

Human designers, by contrast, contribute capabilities that remain difficult to replicate computationally, including: contextual reasoning; empathy with users; ethical judgement; aesthetic evaluation.

Research in human–computer interaction suggests that effective human–AI collaboration involves iterative feedback loops between computational generation and human evaluation (Amershi et al., 2019). In these workflows, designers prompt AI systems to generate ideas, evaluate the outputs, refine their prompts, and repeat the process until meaningful solutions emerge.

This collaborative process positions AI as a cognitive amplifier, expanding the search space of ideas while leaving the responsibility for interpretation and decision-making with human designers (Willems et al., 2024). Within educational contexts, scholars have also explored how AI systems can transition from being used purely as tools toward acting as collaborative partners in learning processes (Sockalingam et al., 2025). This shift from AI as tool to AI as teammate highlights the importance of teaching students how to critically interpret AI outputs rather than simply generating them.

However, the increasing accessibility of generative AI tools also raises concerns about the potential for cognitive outsourcing, where individuals rely excessively on automated systems rather than developing their own analytical and creative capabilities (Khan et al., 2025).

**2.3 AI Integration in Design Education**

The introduction of generative AI tools into educational environments presents both opportunities and challenges for design pedagogy. On one hand, AI systems can support learning by enabling students to explore ideas rapidly, synthesize large amounts of information, and experiment with alternative design directions. Such capabilities can accelerate early stages of the design process and expose students to a wider range of conceptual possibilities.

On the other hand, educators have expressed concern that easy access to AI-generated outputs may reduce opportunities for students to develop fundamental design skills such as observation, critical analysis, and creative reasoning (Khan et al., 2025). If students rely on AI to generate ideas or interpret research findings without engaging deeply with the design context, the educational value of design projects may be diminished.

Studio-based design education has traditionally emphasized experiential learning, where students develop insights through field observation, prototyping, and iterative experimentation (Schön, 1983). These activities encourage reflective practice and enable learners to build contextual understanding of complex design problems.

Recent scholarship argues that the integration of AI into design education should not focus solely on teaching students how to use AI tools. Instead, educators should prioritize the development of higher-order cognitive capabilities that allow students to work critically and creatively with AI systems (Huang et al., 2025).

Within this emerging pedagogical perspective, the goal of design education shifts toward cultivating the human skills necessary for effective human–AI collaboration.

**Theoretical Framework: SuperSkillsStack**

To analyze how students engage with AI tools during design projects, this study adopts the SuperSkillsStack framework, which conceptualizes the human capabilities required for meaningful collaboration with AI systems (Huang et al., 2025).

The framework proposes that effective human–AI collaboration depends on the development of four interconnected capabilities: Agency; Domain Knowledge; Imagination and Taste.

Together, these capabilities form a cognitive structure that allows individuals to harness the benefits of AI while maintaining human responsibility for interpretation and decision-making.

1) **Agency**

Agency refers to the ability of individuals to take initiative and exert control over technological systems. In the context of AI-assisted design, agency involves deciding how and when AI tools should be used within the design process.

Designers with strong agency do not passively accept AI-generated outputs. Instead, they actively:

- experiment with different tools
- refine prompts to improve outputs
- determine which stages of the design process benefit from AI assistance

Agency therefore reflects the ability to orchestrate human–AI collaboration, positioning AI systems as instruments within a broader design workflow rather than autonomous decision-makers (Willems et al., 2024).

The concept of agency is particularly important in educational contexts, where students must learn not only how to operate AI tools but also how to critically assess their usefulness.

**2) Domain Knowledge**

Domain knowledge refers to the contextual understanding required to interpret design problems and evaluate proposed solutions.

In design practice, domain knowledge emerges through activities such as: observing real environments; conducting user interviews; analyzing spatial and social contexts; understanding cultural practices.

Because generative AI systems rely on generalized training data rather than direct experience, they often produce outputs that lack the specificity required for particular design contexts (Floridi & Chiriatti, 2020).

Domain knowledge therefore enables designers to identify inaccuracies or oversimplifications in AI-generated suggestions. Designers can determine whether AI-generated ideas align with real-world observations or whether they require modification.

In this sense, domain knowledge functions as a critical filter that ensures AI-generated outputs are grounded in the realities of the design context.

**3) Imagination**

Imagination refers to the capacity to generate novel ideas and envision alternative possibilities.

In design thinking, imagination is closely associated with the divergent phase of the creative process, where designers explore multiple potential solutions before narrowing down to specific concepts (Brown, 2008).

Generative AI systems can support imagination by producing a wide range of design variations quickly. These variations can inspire new directions or reveal unexpected connections between ideas.

However, imagination is not simply the generation of ideas. It also involves interpreting, combining, and developing those ideas into coherent design concepts. Human designers play a central role in this process by selecting and elaborating on ideas that align with the goals of the design project.

Thus, imagination emerges through an interaction between human creativity and computational generation.

**4) Taste**

Taste refers to the ability to evaluate the aesthetic, conceptual, and practical quality of ideas.

In creative disciplines, taste involves recognizing which ideas are meaningful, feasible, and appropriate within a given context. Taste is developed through experience and exposure to examples of effective design.

Because generative AI systems can produce large quantities of ideas quickly, the ability to evaluate and refine those ideas becomes increasingly important. Designers must decide which outputs are worth pursuing and which should be discarded.

The concept of taste echoes broader discussions of aesthetic judgement in creative practice, where the ability to discern quality is often considered a hallmark of expertise (Boden, 2016). Recent scholarship on the future of expertise argues that emerging technologies require individuals to develop multifaceted mastery, combining technical knowledge with interpretive and creative capabilities (Poon, Willems, & Liu, 2023).

Within AI-assisted workflows, taste functions as the final evaluative stage of the design process, guiding the selection and refinement of ideas generated through human–AI collaboration.

**5) Integrating the SuperSkillsStack**

The four capabilities of the SuperSkillsStack framework operate together as an integrated system.

- Agency determines how AI tools are deployed
- Domain Knowledge ensures contextual accuracy
- Imagination expands the range of possible ideas
- Taste evaluates and refines design outcomes

Together, these capabilities enable designers to harness AI technologies without relinquishing control over the creative process.

In design education, cultivating these capabilities may help students navigate the rapidly evolving landscape of generative AI while maintaining the human-centered foundations of design practice.

**Methodology**

**1) Research Design**

This study employed a qualitative thematic analysis to examine how students integrated generative AI tools into their design projects. The analysis focused on reflective writings produced by student teams at the end of a design-based learning assignment.

Reflective writing was chosen as the primary data source because it provides insight into students' reasoning, decision-making processes, and perceptions of AI-assisted learning. Unlike observational data alone, reflections capture how learners interpret their own design processes and evaluate the role of technology within them.

The analytical framework for this study was derived from the SuperSkillsStack model, which identifies four key human capabilities necessary for effective collaboration with AI systems: Agency, Domain Knowledge, Imagination, and Taste.

These four dimensions served as the primary coding categories for analyzing the reflections.

**2) Educational Context**

The reflections analyzed in this study were produced as part of a design thinking and innovation course in which students conducted human-centered design projects focused on real-world environments.

Each student team was required to: Conduct site analysis of a selected location; Perform user research through interviews or observations; Develop problem statements and design insights; Produce visual documentation including diagrams, journey maps, and presentations; Reflect on their use of AI tools during the design process.

Students were encouraged to experiment with generative AI systems such as:ChatGPT, Gemini, Perplexity, Claude, Microsoft Transcribe and Image-generation tools.

However, AI usage was not mandatory. Students were free to decide how and when AI would be incorporated into their workflow.

This open-ended approach allowed researchers to observe naturally emerging patterns of human–AI collaboration in the design process.

3) **Dataset Description**

The dataset consisted of reflective submissions from 80 student teams.

Each team reflection addressed several prompts related to AI integration, including: Moments where AI improved the team's design learning; Situations where AI did not work well; Actions taken to address AI limitations; Decisions about which AI tools to use; Examples of AI used as a tool, teammate, or not used at all; Whether AI supported creative thinking. The reflections varied in length but generally contained qualitative descriptions of the design process and team decision-making.

Across the dataset, students described working on a diverse range of design contexts including: public transport environments; hawker centres; healthcare clinics; residential spaces; gyms; parks and urban environments, etc.

These varied contexts provided a rich dataset for examining how AI tools interact with different forms of design research and fieldwork.

4) **Data Preparation**

All reflections were compiled into a single dataset and anonymized to remove identifying information. Each team was assigned a group identifier to preserve the structure of the reflections while protecting participant identity.

The reflections were then segmented into meaningful analytical units, typically consisting of individual sentences or short paragraphs describing AI usage or decision-making.

These units served as the basis for qualitative coding.

5) **Coding Procedure**

The analysis followed a two-stage coding process combining deductive and inductive approaches.

Stage 1: Framework-Based Coding

In the first stage, all reflection segments were coded according to the four SuperSkillsStack categories:

| Code | Description |
| --- | --- |
| Agency | Students demonstrating initiative in deciding how AI was used |
| Domain Knowledge | Students using contextual understanding to verify AI outputs |
| Imagination | AI supporting idea generation or brainstorming |
| Taste | Students evaluating and refining AI-generated outputs |

Each segment of text could receive multiple codes if it reflected more than one capability.

For example, a statement describing the rejection of an AI-generated diagram based on site observations could be coded as both Domain Knowledge and Taste.

Stage 2: Thematic Pattern Analysis

After initial coding, the researchers conducted a second round of analysis to identify broader patterns in the data.

This stage involved examining how the coded segments clustered around specific stages of the design process, including: site observation, ideation, analysis, evaluation, refinement.

Through this process, a human–AI design workflow model emerged, illustrating how AI was primarily used in early stages of ideation while human judgement dominated later stages of interpretation and evaluation.

6) **Data Analysis**

Following the coding process, the frequency of each SuperSkillsStack capability was calculated across the dataset. This quantitative summary complemented the qualitative analysis and helped identify the relative prominence of different capabilities in the reflections.

The analysis revealed that: Taste appeared in every group reflection; Domain Knowledge appeared in most reflections; Agency appeared frequently in descriptions of AI tool selection; Imagination appeared primarily in early-stage brainstorming activities.These findings informed the thematic structure of the Results section.

**Results**

Analysis of the student reflections revealed consistent patterns in how learners interacted with AI tools during the design process. Using the SuperSkillsStack framework—Agency, Domain Knowledge, Imagination, and Taste: four major themes emerged from the data. These themes illustrate how students positioned AI within their design workflow and how human judgement remained central to the learning process.

Across groups, students used AI most frequently during early ideation and information organization. However, they consistently relied on human judgement, contextual understanding, and real-world observation to interpret findings and refine solutions. The findings below describe how each SuperSkillsStack capability manifested in student reflections.

**1) Agency: Students Exercising Control Over AI Use**

Agency was evident in the ways students actively decided when and how to integrate AI tools into their workflow. Rather than relying on AI automatically, many groups described making deliberate decisions about the appropriate role of AI within their design process. Students frequently positioned AI as a supporting tool rather than a primary decision-maker.

A common pattern involved restricting AI use during stages requiring direct observation or human interpretation. Students emphasized that design insights must emerge from engagement with real environments and users rather than AI-generated assumptions.

Students wrote:

> *"AI was not used for the site analysis at all as we felt that generating observations using AI could not reflect the physical conditions of the site."*

Another group similarly highlighted the limitations of AI during ethnographic activities.

> *"We avoided AI during user interviews and empathy mapping because AI cannot interpret body language, tone, or user discomfort."*

Some teams also described deliberately stopping AI use when they noticed that generated outputs were misleading or overly generic.

> *"We completely stopped using AI for the interpretation of qualitative data because it ignored important observations from our fieldwork."*

Agency was also demonstrated in students' selection and experimentation with different AI **platforms. Rather than relying on a single system, students compared outputs across** multiple tools to determine which were most useful for specific tasks.

One group explained:

> *"We experimented with different platforms and found Perplexity and Gemini to give us the best output after trial and error."*

Another group emphasized the importance of accessibility and iterative prompting in their choice of tools.

> *"We chose ChatGPT because everyone in the team had access to it and it allowed quick back-and-forth prompting."*

These reflections indicate that students did not treat AI as an authoritative source of knowledge. Instead, they actively **managed and orchestrated AI use**, demonstrating strong agency in determining how technology contributed to their design process.

2) **Domain Knowledge: Validating AI Through Real-World Context**

Domain knowledge emerged as a crucial factor enabling students to evaluate and interpret AI-generated outputs. Many reflections described situations in which AI produced plausible but inaccurate information about the design context. Students relied on site visits, interviews, and observational data to verify and correct these outputs.

Several groups reported that AI-generated diagrams or site descriptions did not match the physical conditions observed during fieldwork.

One reflection explained:

> *"The generated images didn't match the site layout or the visual style we needed, which made the results unusable for our project."*

Another group similarly noted discrepancies between AI-generated information and their on-site observations.

> *"AI sometimes generated convincing but inaccurate insights that didn't match our site data."*

Students repeatedly emphasized that contextual understanding emerged only through physical presence in the environment.

One team described the value of experiential knowledge gained through fieldwork:

> *"When we were physically there, we could actually feel the environment and relate to the users' frustrations, which helped us see the problem more clearly."*

Another group observed that AI lacks the experiential understanding necessary for site-specific analysis.

> *"AI does not have experience visiting the site nor interacting with the people there."*

These reflections demonstrate that domain knowledge functions as a **critical filter for evaluating AI outputs**. Students relied on their contextual understanding of environments and user behaviors to determine which AI-generated insights were relevant and which were misleading.

### 3) Imagination: AI as a Catalyst for Idea Generation

Students most frequently described AI as beneficial during the early stages of ideation. Many groups reported that AI helped them overcome creative blocks and explore alternative perspectives during brainstorming.

Several reflections highlighted AI's ability to generate initial ideas quickly.

One group wrote:

> *"AI was helpful for creative thinking as we were able to use it to make the process of initial ideation much easier and faster."*

Another group described how AI expanded the range of possibilities considered during brainstorming.

> *"AI strongly supported divergent thinking by expanding our idea pool and offering fresh angles quickly."*

Students also reported using AI to refine problem statements and explore multiple variations of design questions.

One reflection described how reviewing different AI-generated versions helped clarify the team's design direction.

> *"Reviewing these helped us identify what our core intention really was and solidified the direction of our analysis."*

Similarly, another group described AI as a conversational partner that helped stimulate discussion within the team.

> *"Even though the suggestions were generic, it helped us get started and react to what we agreed or disagreed with."*

However, students consistently emphasized that AI-generated ideas served primarily as starting points rather than final solutions.

One group reflected:

> *"AI gave us a helpful starting point in creative thinking that we could further build and elaborate upon."*

This suggests that AI functioned as a catalyst for imagination, accelerating the exploration of ideas while leaving the development and refinement of concepts to human designers.

4) **Taste: Human Judgement in Evaluating AI Outputs**

Taste, defined as the ability to evaluate the quality and appropriateness of ideas, appeared most prominently during the later stages of the design process. Students frequently described rejecting or modifying AI-generated outputs that did not align with their observations or design goals.

Many reflections noted that AI-generated suggestions often sounded convincing but lacked contextual relevance.

One group explained:

> *"Sometimes AI provided ideas that sounded polished but were too broad or unrealistic for our context."*

Another team described how AI-generated explanations failed to capture the complexity of their research findings.

> *"AI explanations sounded too generic and did not fully match the nuances of our site observations."*

Students responded to these limitations by filtering and refining AI-generated outputs through collaborative discussion and manual revision.

One group described implementing a systematic approach for evaluating AI-generated ideas.

> *"We created a checklist to evaluate suggestions based on feasibility, relevance to interview insights, and alignment with the design brief."*

Another reflection emphasized the importance of rewriting AI-generated text to ensure accuracy and authenticity.

> *"Instead of copying AI text directly, we used it as a starting point and then rewrote the content in our own voice."*

Students also described correcting visual outputs produced by AI systems.

One group wrote:

> *"We manually sketched and refined the maps ourselves when the AI outputs were inaccurate or messy."*

These reflections demonstrate that students relied on human judgement to determine which ideas were meaningful and feasible. Taste therefore functioned as the final evaluative stage of the design process, ensuring that AI-generated outputs were critically assessed and refined.

**5) Synthesis: Human–AI Collaboration in the Design Process**

Taken together, the reflections reveal a consistent pattern in how students integrated AI into their design workflow. AI was primarily used to accelerate early stages of the process, such as brainstorming, summarizing information, and organizing research materials. However, students consistently relied on human judgement to interpret observations, validate findings, and refine design outcomes.

Across the reflections, a recurring workflow emerged:

1. AI generates initial ideas and summaries
2. Students interpret outputs using domain knowledge
3. Teams filter ideas through critical discussion
4. Students refine designs through human judgement

This pattern suggests that AI functioned primarily as a cognitive accelerator, enabling students to explore ideas more rapidly. At the same time, the core activities of design—contextual interpretation, empathy with users, and aesthetic evaluation—remained fundamentally human.

These findings illustrate how the SuperSkillsStack capabilities operate collectively in design education. Agency governs how AI tools are used, domain knowledge ensures contextual accuracy, imagination expands the range of possible ideas, and taste determines which ideas ultimately contribute to meaningful design solutions.

**Discussion**

This study examined how students integrated generative AI into their design projects through the lens of the SuperSkillsStack framework: Agency, Domain Knowledge, Imagination, and Taste. The findings demonstrate that while AI tools played an important role in accelerating certain stages of the design process, students consistently relied on human capabilities to interpret context, evaluate ideas, and produce meaningful design outcomes.

Three major insights emerge from the analysis:
(1) AI functions primarily as a cognitive accelerator for exploration,
(2) domain knowledge and taste act as safeguards against AI limitations, and
(3) design education can use AI integration to strengthen rather than weaken human capabilities.

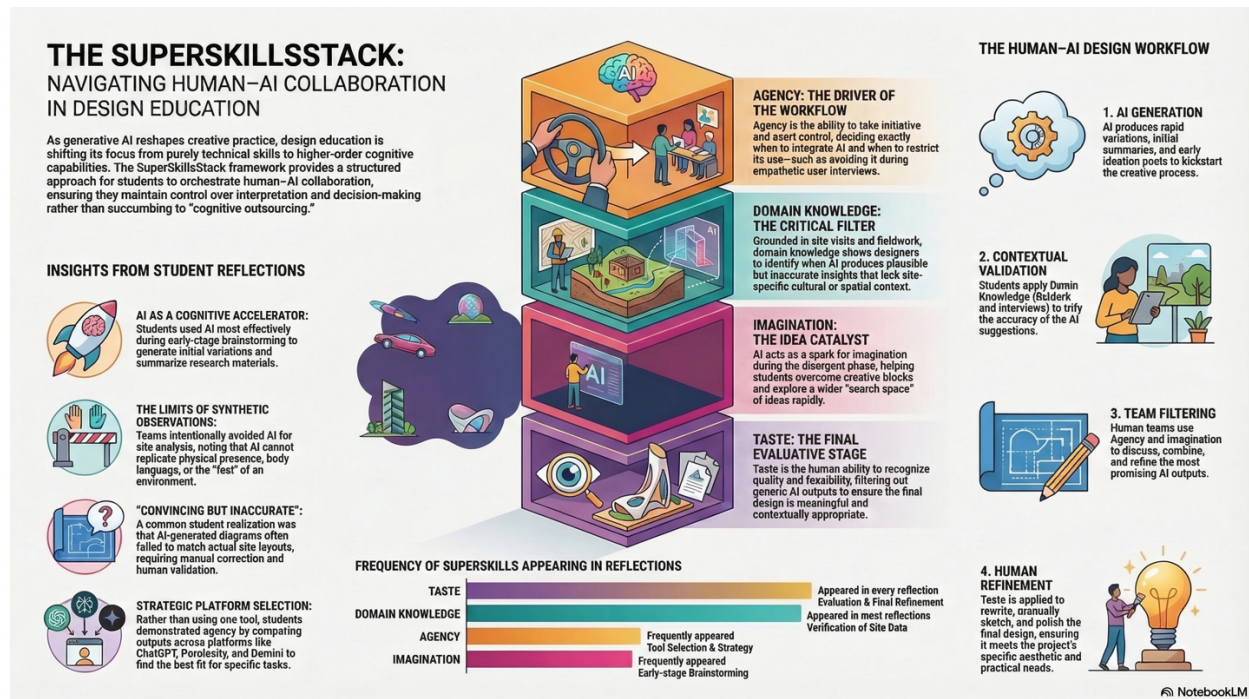

Figure 1. The "Trilingual Triad" generated by NotebookLM

### 1) AI as a Cognitive Accelerator for Early Design Stages

One of the most consistent findings across the student reflections is that AI tools significantly improved the efficiency of early-stage design activities. Students frequently used AI to generate initial ideas, summarize research notes, organize information, and refine problem statements. These activities correspond to the divergent phase of the design process, where teams explore multiple possibilities before narrowing down to specific solutions.

Students described AI as helpful in overcoming creative inertia, particularly when starting a new project or framing design problems. The ability to rapidly generate variations of ideas allowed teams to move quickly from vague concepts to more structured problem statements.

However, while AI accelerated idea generation, students did not treat AI-generated suggestions as final answers. Instead, they used them as starting points that required further interpretation and development. This indicates that AI primarily supported exploratory thinking, while the deeper processes of design reasoning remained human-driven.

This pattern aligns with broader research on human–AI collaboration, which suggests that generative AI is particularly effective at expanding the search space of ideas but less capable of evaluating their contextual relevance.

**2) Domain Knowledge as a Critical Filter for AI Outputs**

Another key finding is the importance of domain knowledge in evaluating AI-generated outputs. Students frequently reported that AI produced plausible but inaccurate information about their design sites, including incorrect diagrams, generic insights, and misinterpretations of user behavior.

These limitations were most apparent when AI attempted to interpret site-specific contexts such as spatial layouts, cultural practices, or human interactions within environments. Because AI systems rely primarily on generalized training data rather than direct observation, their outputs often lacked the specificity required for design analysis.

Students therefore relied heavily on fieldwork activities such as site visits, interviews, and observational research to validate AI outputs. These activities allowed them to identify discrepancies between AI-generated suggestions and real-world conditions.

The findings highlight the continued importance of experiential learning in design education. While AI tools can provide useful starting points, they cannot replace the contextual understanding gained through direct engagement with environments and users.

**3) Taste and Human Judgement in Design Evaluation**

Perhaps the most prominent capability observed in the reflections was taste**,** defined as the ability to evaluate the quality, relevance, and feasibility of design ideas.

Students consistently described situations where AI-generated suggestions appeared convincing but were ultimately unsuitable for their design contexts. In response, teams engaged in critical evaluation processes that involved rewriting text, redesigning diagrams, and filtering ideas based on feasibility and alignment with research findings.

This evaluative process reflects a fundamental characteristic of design practice: the ability to determine which ideas are meaningful within a given context. AI systems may generate large quantities of ideas, but determining which of those ideas are valuable remains a distinctly human capability.

The prominence of taste across all reflections suggests that evaluative judgement is a core skill that design education should continue to emphasize, particularly in environments where AI-generated content is readily available.

**4) Agency in Human–AI Collaboration**

The reflections also demonstrate that students exercised considerable agency in determining how AI tools were used during their projects. Rather than passively accepting AI outputs, students actively decided when AI should be integrated into the workflow and when it should be excluded.

Many groups described intentionally avoiding AI during stages of the design process that required empathy, contextual interpretation, or human observation. Students also experimented with multiple AI tools and refined their prompts to improve the relevance of generated outputs.

This pattern suggests that effective human–AI collaboration depends not only on technical literacy but also on the ability to strategically deploy AI tools within the broader design process. Agency therefore becomes a key capability that enables students to use AI productively without becoming dependent on it.

**5) Implications for Design Education**

The findings of this study offer several implications for design education in the era of generative AI.

First, AI should be introduced as a collaborative design partner rather than a replacement for creative thinking. Students should be encouraged to use AI tools to accelerate exploration and experimentation while maintaining responsibility for interpretation and evaluation.

Second, design curricula should continue to emphasize field research and contextual learning. Site visits, interviews, and observational methods remain essential for developing the domain knowledge required to assess AI-generated insights.

Third, educators should focus on cultivating design judgement and critical evaluation skills. As AI tools make it easier to generate ideas and artifacts, the ability to distinguish meaningful solutions from superficial ones becomes increasingly important.

Finally, AI literacy should be integrated into design education not only as a technical skill but also as a strategic competency. Students must learn how to decide when AI is useful, when it is misleading, and how to combine AI outputs with human insight.

**Conclusion**

This study examined how students integrated generative AI tools into their design projects through an analysis of reflective writings from multiple design teams. Using the "SuperSkillsStack framework—Agency, Domain Knowledge, Imagination, and Taste",the study identified key patterns in how students collaborated with AI during the design process.

The findings indicate that AI plays a valuable role in accelerating early-stage design activities such as brainstorming, information synthesis, and problem framing. However, the deeper processes of design—including contextual interpretation, empathy with users, and critical evaluation of ideas—remain fundamentally human.

Students demonstrated strong agency by deliberately deciding when and how AI tools should be used. Domain knowledge gained through fieldwork allowed them to validate and correct AI-generated outputs. Imagination was supported by AI during ideation, enabling teams to explore a wider range of possibilities. Most importantly, taste guided the evaluation and refinement of ideas, ensuring that design outcomes remained grounded in real-world contexts.

Taken together, these findings suggest that generative AI does not replace the human capabilities required for design thinking. Instead, it reshapes the design process by accelerating exploratory stages while increasing the importance of human judgement and contextual understanding.

For design education, this shift highlights the need to cultivate the human capabilities that enable meaningful human–AI collaboration. By developing agency, domain knowledge, imagination, and taste, students can learn to leverage AI tools effectively while preserving the critical thinking and creative judgement that define design practice.

Future research could extend this study by examining how these capabilities develop over time as students gain more experience working with AI tools, or by comparing learning outcomes between AI-assisted and traditional design studio environments. Such research will be essential for understanding how design education can continue to evolve in response to rapidly advancing AI technologies.